\documentclass[11pt,a4paper]{article}
\usepackage{epsfig}
\begin{document}
\title{Disjoining pressure of planar adsorbed films}
\author{J R Henderson\\
School of Physics and Astronomy, University of Leeds, Leeds LS2 9JT, UK}
\date{\today}
\maketitle
\begin{abstract}
\noindent Frumkin-Derjaguin theory of interfacial phase transitions and in particular the concept of the disjoining pressure of a planar adsorbed film is reviewed and then discussed in terms of statistical mechanical formulations of interfacial phase transitions beyond mean-field.
\end{abstract}
\newpage
\section*{1.\,\, Frumkin-Derjaguin Theory of Wetting Phenomena}
The thermodynamics of interfacial phase transitions (IPT) was founded over 70 years ago by 
Frumkin \cite{frumkin} and combined with the concept of the `disjoining pressure' introduced even earlier 
by Derjaguin in the same context \cite{frumderj} and in the context of adsorption in porous media \cite{derj1,derj2}. 
Frumkin was directly concerned with the stability of a film of liquid adsorbed at a planar solid-gas interface, which he treated as the direct generalisation of van der Waals's molecular theory of bulk phase transitions to interfacial phase transitions. In particular, he considered the thermodynamic stability of a planar fluid film as a function of 
its thickness by implicitly assuming that over-saturated gas can exist in 
the absence of bulk liquid whenever required to enable the system to traverse 
a van der Waals loop within, for example, an adsorption isotherm. The interfacial thermodynamic free-energy is the surface-excess grand potential of the liquid film per unit area, $\gamma_{SG}$, historically known as surface tension. When the bulk free energy terms are subtracted via appropriate choices of Gibbs dividing surfaces, the conjugate pair of thermodynamic variables for isothermal processes (which in bulk are chemical potential $\mu$ and number density $\rho$) translate to (field, density) = ($\mu$, $\Gamma$), where $\Gamma$ denotes the adsorption (excess amount of molecules adsorbed per unit area). In the context of wetting films, as considered by Frumkin, it is natural to use a suitably defined film thickness $\ell$ (rather than $\Gamma$) as the effective order parameter, which can be done rigorously by defining the thickness to be the adsorption divided by a number density that is fixed for a given isotherm; the obvious choice suitable for all but unphysically thick unsaturated films is $\ell \equiv \Gamma/\Delta\rho$ where for any particular isotherm $\Delta\rho \equiv \rho_L-\rho_V$ is the difference between the saturated liquid density and the density of coexisting vapour at the chosen temperature $T$. A transformation from the field $\mu$ to the order parameter $\ell$
is then via the usual transformation between the grand ensemble and the canonical ensemble:
\begin{equation}
\gamma_{SG}(\mu) = f(\ell)-\mu\,\Delta\rho\,\ell\,\,,
\label{trans}
\end{equation}
where $f(\ell)$ denotes the interfacial excess Helmholtz free energy per unit area. This 
statistical-thermodynamics immediately yields the second-law of thermodynamics in two 
equivalent forms:
\begin{eqnarray}
\partial\gamma_{SG}(\mu) &=& -\Delta\rho\,\,\ell\,\partial\mu \label{2ndmu}\\
\partial f(\ell) &=& \mu\,\Delta\rho\,\partial\ell\,\,.
\label{2ndell}
\end{eqnarray}
The first of these results (\ref{2ndmu}) is just the Gibbs adsorption equation. Frumkin and Derjaguin use the mean-field quantity $\mu(\ell)$ to define a disjoining pressure 
\begin{equation}
\Pi(\ell) \equiv \left[\mu_o-\mu(\ell)\right]\Delta\rho\,\,,
\label{pimu}
\end{equation}
where $\mu_o$ denotes the chemical potential corresponding to an equilibrium film thickness $\ell_{o}$.
Note that by construction one can read off the disjoining pressure, for any chosen $\mu_o$, directly from 
an experimental adsorption isotherm (i.e. $\ell$ versus $\mu$) and hence we can hereafter write $\mu(\ell)$ as introduced in Eq.~(\ref{pimu}). It is also possible to introduce a mean-field potential $\psi(\ell)$, an example of what we now call a Landau potential or in this specific context the interface potential, by imagining a constraint sufficient to stabilize planar adsorbed films away from equilibrium:
\begin{eqnarray}
\psi(\ell;\mu_o) &\equiv& f(\ell)-\mu_o\,\Delta\rho\,\ell\,\,,\label{vee}\\
\Pi(\ell) &=& -\frac{\partial \psi(\ell)}{\partial\ell}\,\,.
\label{dispot}
\end{eqnarray}
Once a specific $\mu_o$ has been chosen, all values of the film thickness except that for which $\mu(\ell)=\mu_o$ represent the free-energy of metastable or unstable planar films. At $\mu(\ell)=\mu_o$ there is always at least one value of $\ell$ corresponding to the absolute minimum of $\psi(\ell)$, where the interface 
potential equals the equilibrium surface tension $\gamma_{SG}(\mu_o)$. \\

The Russian School summarise the above statistical-thermodynamics with the following transformation of the second law of interfacial thermodynamics:
\begin{equation}
\gamma_{SG} = \psi + \ell\Pi\,;\,\,\,\,\partial \psi = -\Pi\partial\ell\,;\,\,\,\,\partial\gamma_{SG} 
= \ell\partial\Pi\,\,.
\label{russ}
\end{equation}
Here one can note three distinct but equivalent approaches to applying Frumkin-Derjaguin theory in practice. Frumkin emphasised 
plots of the thermodynamic potential versus the inverse of the order parameter; i.e. $\gamma_{SG}$ versus $1/\ell$ which is the direct equivalent of a $p-V$ isotherm in van der Waals' theory of bulk phase transitions. Derjaguin preferred to analyse the field as a function of its conjugate order parameter; i.e. $\mu(\ell)$ or equivalently adsorption isotherms $\ell(\mu)$. 
In either case, a first-order phase transition between equilibrium films of different thickness $\ell_{o}$ and $\ell_{1}$ would appear as a van der Waals loop associated with an equal-area Maxwell construction: 
\begin{eqnarray}
\int_{1/\ell_{1}}^{1/\ell_{o}}d\frac{1}{\ell}\gamma_{SG}(\ell) &=& \gamma_{SG}(\mu_o)\left(\frac{1}{\ell_{o}}-\frac{1}{\ell_{1}}\right)\,\,,
\label{frumderj2}\\
\int_{\ell_{o}}^{\ell_{1}}d\ell\Pi(\ell) &=& 0\,\,,
\label{fderj3}
\end{eqnarray}
respectively \cite{footeq}. The final route, based on identifying $\psi(\ell)$, was adopted by physicists as the work-horse of density functional theory (DFT) of inhomogeneous fluids \cite{evansrev}, although the nature and significance of the constraint needed to stabilise a film out of equilibrium has typically been treated rather vaguely \cite{jrhdis}. In the vicinity of a 
first-order interfacial phase transition $\psi(\ell)$ develops a second minimum directly related to the Maxwell construction for a van der Waals loop in $\Pi(\ell)$: 
\begin{equation}
\psi(\ell_{o})=\psi(\ell_{1})=\gamma_{SG}(\mu_o)\,\,.
\label{landth}
\end{equation}
In the absence of a first-order phase transition in the approach to saturation ($\mu = \mu_{sat}$) every chosen value of
$\mu_o < \mu_{sat}$ corresponds exclusively to an equilibrium system ($\ell_{o} = \ell(\mu=\mu_o$)). Then when the above statistical thermodynamics is applied to partial wetting at saturation (defining an equilibrium planar film surrounding an adsorbed drop of excess liquid with a non-zero value of Young's contact angle) 
one obtains probably the most well-known expression from Frumkin-Derjaguin 
theory \cite{derj1}
\begin{equation}
\int_{\ell_{o}}^{\infty}d\ell\Pi(\ell) = \psi(\ell_{o})-\psi(\infty) = \gamma_{LV}(\cos\theta -1)\,\,,
\label{frumderj}
\end{equation}
where the final equality has introduced Young's equation \cite{young} for the contact angle 
$\theta$. The above result is a direct expression of the thermodynamic work needed to quasi-statically thin a 
macroscopically thick film of excess liquid down to its equilibrium thickness. \\

Frumkin-Derjaguin theory was used as the thermodynamic basis of 
the famous analysis of Dzyaloshinskii, Lifshitz and Pitaevskii (DLP), who 
adapted high-temperature quantum field-theory to directly calculate the dispersion 
interaction contribution to the normal component of the pressure tensor and hence obtain the excess pressure on the surface due to the presence of an adsorbed film. By identifying this quantity with Derjaguin's disjoining pressure DLP were able for the first time to show that one could explicitly implement Frumkin-Derjaguin theory of interfacial phase transitions within theoretical physics \cite{DLP1,DLP2}. This work was adopted as a foundation stone of modern colloid and interface science, despite the clear warning of DLP that their field-theory could only be trusted for macroscopically thick films (films surrounding drops with a Young's contact angle of no more than about 0.1 degrees \cite{DLP2}). Much harder to explain is why 
it was more than 15 years after DLP before the interfacial phase transitions of Frumkin were `rediscovered' and named wetting transitions \cite{cahn,ebsaam,Nak-Fish}. Direct use of Frumkin-Derjaguin theory constitutes a mean-field theory of interfacial phase transitions (MFIPT), because of the constraint required to define a van der Waals loop through values of the film thickness where a planar film would be metastable or unstable to long-wavelength capillary wave fluctuations (eventually broadening into coexistence between films of different thicknesses). Interestingly, the upper critical dimension at and below which $\psi(\ell)$ is renormalised by long-wavelength capillary-wave fluctuations away from planarity is known to be $d=3$ for 
short-ranged models \cite{lip83} and in the presence of dispersion forces naturally occurring IPT are believed to be mean-field \cite{lip84}. \\

\section*{2. Physical interpretation of the mean-field disjoining pressure}
In his monograph, Frenkel \cite{Frenkel} reports that Derjaguin's disjoining pressure was apparently first discovered ``by D. Talmud and S. Bresler, who noticed in 1931 that two drops of mercury freely moving on the bottom of a glass vessel filled with molten paraffin did not coalesce but remained separated from each other by a very thin paraffin film'', \cite{Bresler}. Frenkel chooses to review this work in the context of the ``specific elasticity of thin films''. However, the name ``disjoining pressure'' indicates that the Russian School understood that Eq.~(\ref{pimu}) must represent the force per unit area exerted by the fluid film on the substrate, minus that which would be present for the equilibrium film at $\mu=\mu_o$. In fact, DLP  appear to assume that the normal component of the pressure tensor is directly proportional to the film chemical potential \cite{DLP1}. In which case one could envisage direct measurements of the disjoining pressure by placing pressure sensors just behind the substrate surface. Mean-field DFT has been used to successfully test this deep connection within statistical mechanics \cite{jrhdis}. In particular, for a wall-fluid model in which the substrate field is represented as a one-body external field $v^{ext}(z)$ the statistical mechanical definition of the disjoining pressure must be
\begin{equation}
\Pi(\ell) = -\int_{-\infty}^{\infty}dz\left[\rho(z;\ell)-\rho(z;\ell_o)\right]v^{ext\,'}(z)\,\,,
\label{pism}
\end{equation}
where $\rho(z;\ell_o)$ is the density profile of the adsorbed film at equilibrium (at chemical potential $\mu_o$). At saturation ($\mu_o=\mu_{sat}$) one can replace $\rho(z;\ell_o)$ with the wall-liquid density profile $\rho_{SL}(z)$, for the purpose of calculating the disjoining pressure from Eq.~(\ref{pism}).  DLP express Eq.~(\ref{pism}) in terms of the normal component of a pressure tensor. The full tensor can be defined to within an arbitrary path integral (or gauge) from the many-body expression for the total momentum $\dot{J}({\mbox{\bf r}},t)$ \cite{hensch}. The conservation of momentum $\langle\dot{J}({\mbox{\bf r}},t)\rangle=0$ leads to a sum rule for the equilibrium pressure tensor:  
\begin{equation}
\nabla^{\beta}p^{\alpha\beta}({\mbox{\bf r}})=-\rho({\mbox{\bf r}})\nabla^{\alpha}v^{ext}({\mbox{\bf r}})\,\,.
\label{pten}
\end{equation}
In planar symmetry this reduces to a sum rule for the normal component of the pressure tensor, as required to justify Eq.~(\ref{pism}):
\begin{equation}
p_{N}\,'(z) = -\rho(z)v^{ext\,'}(z)\,\,.
\label{pn}
\end{equation}
The above can be regarded as an expression of Newton's third law; i.e. the force of the fluid on the wall is equal and opposite to that of the wall on the fluid. Thus, there are two equivalent ways of expressing this force balance, either in terms of the intermolecular forces (known as a virial theorem) or, equivalently, in terms of the wall-fluid force as in Eq.~(\ref{pism}), \cite{jrhrev}. \\
\begin{figure}[tbh]
  \center
	\includegraphics[width=14cm,height=8cm,angle=0]{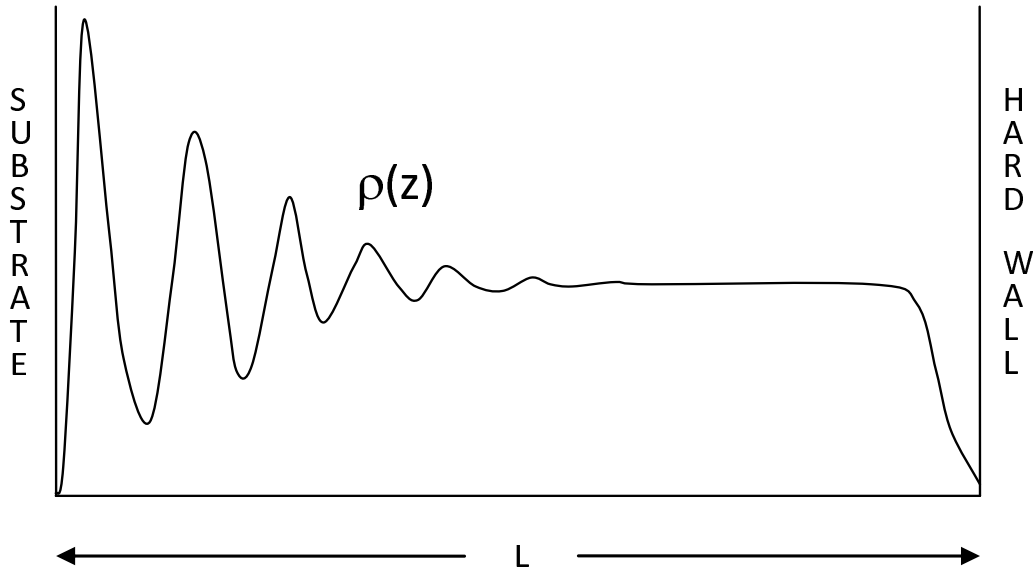}
	\caption{Planar mean-field interfacial theory defined from confinement of a liquid film in an asymmetric planar slit (see text).} 
\end{figure}

A potential illustration of the connection between the statistical mechanical disjoining pressure and the mean-field film chemical potential is sketched in Fig.~1. Here a liquid film is depicted confined between the substrate of interest and a hard-wall boundary. If sum rule (\ref{pn}) is integrated over the hard-wall boundary one finds a conceptually important result:
\begin{equation}
p_{N}(L) = kT\rho_{HW}(L)\,\,,
\label{pnhw}
\end{equation}
where $k$ denotes Boltzmann's constant and the notation highlights the fact that the right-side of (\ref{pnhw}) only applies at the hard-wall boundary; i.e. $\rho(z=L_-) \equiv \rho_{HW}(L)$. If even a nm-thin film of liquid is adsorbed at a substrate then one must be close to saturation and thus one will be able to gently confine the film at a pressure $p_{N}(L)$ close to the vapour pressure. Since the second virial coefficient is negative along the liquid-vapour coexistence curve, one must have a very low value for the density at the hard-wall (provided of course that the substrate field has a range smaller than $L$). Namely, an incipient drying layer must be present at the hard-wall \cite{jrhvs}, as sketched in Fig.~1. In this manner the hard-wall is able to suppress capillary-wave fluctuations away from planar symmetry, without significantly altering the thermodynamic state of the liquid film. In effect, the pore width $L$ is being used to set the thickness $\ell$ of a constrained liquid film. Now consider an infinitesimal change of pore width induced by moving one of boundary walls while keeping the other fixed. If the wall that is moved is the hard-wall, then standard analysis yields \cite{jrhrev,jrhsolv}
\begin{eqnarray}
-\left(\frac{\partial\Omega}{\partial L}\right)_{T,\mu_o} &=& -\int d^3r \rho(z)\frac{\partial v_{HW}(L-z)}{\partial L} \label{solv1}\\
&=& p_{N}(L)A\,\,,
\label{solvf}
\end{eqnarray}
where the integral is over all space (i.e. extends across the hard-wall of area $A$) and sum rule (\ref{pnhw}) has been inserted to obtain the second line. If the substrate wall was moved instead, then a result of identical form to (\ref{solv1}) is obtained, except that the hard-wall field $v_{HW}$ is replaced by the substrate field (note that the sensible way to carry out this alternative derivation is to swap the direction of the $z$ axis so that the substrate is defined to lie at $z=L$ until the final result can be re-expressed back in the geometry of Fig.~1):
\begin{equation}
-\left(\frac{\partial\Omega}{\partial L}\right)_{T,\mu_o} = -A\int_{-\infty}^{L}\rho(z)v^{ext\,'}(z)\,\,.
\label{pnsub}
\end{equation}
Inserting Eq.~(\ref{pn}) then gives back the previous result (\ref{solvf}) since one need only integrate out to the hard-wall or to when the substrate field has fallen to zero (whichever is the least). By defining $\psi \equiv (\Omega + p_oV)/A$, one can rewrite sum rule (\ref{solvf}) in a form directly analogous to (\ref{dispot}):
\begin{equation}
p_{N}(L)-p_o = -\frac{\partial \psi}{\partial L}\,\,,
\label{solvpi}
\end{equation}
except that now the disjoining pressure really is defined in terms of the force per unit area on the hard-wall; $p_o$ is the value $p_{N}(L\rightarrow \infty)$. In this interpretation the disjoining pressure $\Pi(L)$ is the extra pressure on the film needed to maintain a planar film of width $\ell\approx L$ at fixed $T$ and fixed $\mu=\mu_o$. Alternatively, the result can be expressed in terms of the conservation of momentum normal to the substrate wall; i.e. from (\ref{pnsub}). It is this alternative route that is directly expressed in the definition (\ref{pism}). To within the validity of replacing $L$ with $\ell$, we may therefore conclude that the disjoining pressure of Frumkin-Derjaguin theory, defined as a chemical potential difference (\ref{dispot}), is equivalent to the extra pressure on a substrate wall that would be induced by an adsorbed planar film of metastable or unstable thickness. This deep and confusing correspondence is of great significance theoretically, because expressing the disjoining pressure in terms of the conservation of momentum normal to the substrate (\ref{pism}) allows for alternative developments in terms of exact physics \cite{DLP1,mik-weeks,adamjrh}. In principal, experiment has direct access to the local normal force on the substrate wall and so by expressing inhomogeneous fluid phenomena in terms of sum rules involving the exact disjoining pressure, as we shall do in the following section, one would be able to directly analyse experimental systems.

\section*{3.\,\, Beyond mean-field interfacial phenomena}
The physics discussed so far has either implicitly or explicitly assumed the presence of constraints sufficient to prevent large scale fluctuations away from planar symmetry. Thus, any prediction of interfacial phase transitions will be mean-field (MFIPT). To go beyond mean-field one must include these fluctuations. There are standard procedures for doing this in theoretical physics, which essentially boil down to extending the class of constraints to include all physically appropriate classes of symmetry (known as summing over `all' fluctuations). For example, given sufficient numerical resources, one could use DFT to search for solutions of the class $\rho(x,z)$ and $\rho(x,y,z)$, in addition to $\rho(z)$. Frumkin-Derjaguin theory would have to be similarly extended, since Eq.~(\ref{dispot}) implicitly assumes density profiles of the class $\rho(z)$ stabilized by imposing a non-equilibrium chemical potential $\mu(\ell\neq\ell_{o})$. Generalised Frumkin-Derjaguin Theory (GFDT) would need to be based on instantaneous disjoining potential profiles $\hat{\Pi}(x)$ and $\hat{\Pi}(x,y)$, with the sum over fluctuations embodied in formal path-integral definitions defining the full statistical mechanical averages \cite{mik-weeks}
\begin{eqnarray}
\Pi(\ell_o) &=& \langle\hat{\Pi}\rangle \label{dispothat}\\
\ell_o &=&  \langle\hat{\ell}\rangle \,\,,
\label{lhat}
\end{eqnarray}
where the instantaneous film profile is denoted $\hat{\ell}(x)$ or $\hat{\ell}(x,y)$, depending on the class of broken symmetry embodied in the fluctuations. The physical significance of fluctuation physics to interfacial phase transitions is heavily dependent on geometry. If the overall geometry is planar then three-dimensional systems are not strongly affected by thermal  fluctuations, although for short-range models (such as Ising models) fluctuations do generate higher-order contributions to interfacial critical phenomena needed to fully explain the observed physics \cite{parry}. For quasi two-dimensional systems however, even in overall `planar' geometry, interfacial fluctuations are sufficiently dominant to invalidate MFIPT. Let us  therefore restrict further discussion to the geometry of Fig.~2, while bearing in mind the obvious generalisation to higher-dimensional geometry in special cases.  
\begin{figure}[tbh]
  \center
	\includegraphics[width=14cm,height=4cm,angle=0]{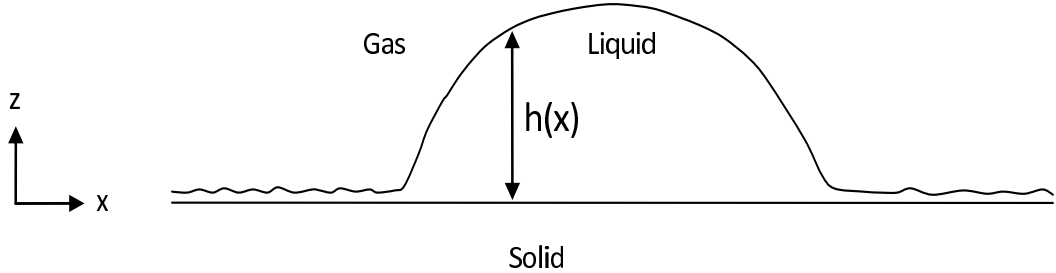}
	\caption{A thermal fluctuation of an adsorbed film away from planar symmetry in quasi two-dimensional geometry. The instantaneous height of the liquid film along the substrate is denoted $h(x)\equiv\hat{\ell}(x)$.} 
\end{figure}
Figure~2 depicts a region of the interface at which a molecularly thin film has fluctuated away from the substrate, known as a `strong' fluctuation regime. Collective fluctuations beyond relevant molecular correlation lengths (typically less than a nm) are much slower than molecular time scales and so one expects local mechanical equilibrium. Thus, it is no accident that the sketch in Fig.~2 reminds readers of Young's picture of an adsorbed drop with a well-defined contact angle arising from the cylindrical (or spherical) cap geometry away from the substrate. Laplace's formulation of local mechanical equilibrium demands that there be an excess pressure inside the drop required to maintain the non-zero curvature $\kappa$ of the film profile. Hence, one can immediately appreciate that a fluctuation of this nature is directly linked to an interfacial pressure wave (surface sound) embodied in the $x$-dependence of the instantaneous disjoining pressure $\hat{\Pi}(x)$. The lowest level of theoretical physics sufficient to describe strong fluctuations requires the replacement of the interface potential $\psi(\ell)$ with an interfacial Hamiltonian functional:
\begin{equation}
H[\hat{\ell}]=\gamma_{LV}\int_{-\infty}^{\infty}dx\left\{\sqrt{1+(\hat{\ell}')^2}-1\right\}+W[\hat{\ell}]\,\,.
\label{iham}
\end{equation}
The first term in the interfacial Hamiltonian defines the cost of distorting the outer surface of the film away from planar symmetry. It might be thought that the use of $\gamma_{LV}$ as the restoring force is problematic for nanoscopic fluctuations but one should note that in this description the film profile $\hat{\ell}(x)$ is already coarse-grained over molecular scales and thus its gradient (the dash denotes $d/dx$) is zero everywhere except where a significant deviation has occurred. The final term in (\ref{iham}) denotes the generalisation of the mean-field interface potential to non-planar geometry; i.e. we must use a non-local functional of $\hat{\ell}(x)$ instead of simply integrating over a mean-field $\psi(\ell(x))$, as emphasised by Parry and coworkers \cite{parry}. $W[\hat{\ell}]$ is known as the binding potential (functional).\\

It is actually possible to use the fundamental statistical mechanics of DFT to regard (\ref{iham}) and hence $W[\hat{\ell}]$ as formally exact functionals defined by the unique relationship between the field $\phi\equiv \mu_o-v^{ext}(x,z)$ and the instantaneous density profile $\rho(x,z)$, \cite{evansrev}. The latter can be used to self-consistently define the instantaneous coarse-grained profile $\hat{\ell}(x)$, as discussed below. That is, $\phi(x,z)$ is the formally exact constraint needed to define an arbitrary film profile. The final result will then be expressed as a formal sum over all physically realisable configurations of the film profile. Note that from this point of view our choice of $\gamma_{LV}$ as the restoring force for bending the film profile is a formal definition, in the same class as choosing a Gibbs dividing surface consistent with macroscopic physics; it effects only the distribution of free-energy between interfacial free energy and higher-order line tension contributions (which it is not particularly informative to include explicitly in the macroscopic Young-Laplace force-balance below). The film profile, for a particular constraint $\phi(x,z)$, is formally given by minimization of the functional $H[\hat{\ell}]$; i.e. $\delta H/\delta \hat{\ell} =0$. The final step is to note that conservation of momentum perpendicular to the substrate requires us to identify the instantaneous disjoining pressure as the functional derivative 
\begin{equation}
\hat{\Pi}[\hat{\ell}] \equiv -\frac{\delta W[\hat{\ell}]}{\delta\hat{\ell}}\,\,,
\label{insdp}
\end{equation}
which in any case is immediately recognisable as the required generalisation of Eq.~(\ref{dispot}) in Frumkin-Derjaguin theory. Namely, with this identification, setting the functional derivative of (\ref{iham}) to zero leads directly to the sum rule
\begin{equation}
\hat{\Pi}(x) = -\gamma_{LV}\frac{d}{dx}\left(\frac{\hat{\ell}'}{\sqrt{1+(\hat{\ell}')^2}}\right) \equiv -\gamma_{LV}\hat{\kappa}(x)\,\,,
\label{dpfin}
\end{equation}
where $\hat{\kappa}(x)$ is the coarse-grained interfacial curvature profile. That this realisation of Laplace's force balance is a direct consequence of the conservation of momentum normal to the substrate can be seen in Fig.~2 if we regard the fluctuation as the formation of a cylindrical-cap section of a drop of radius $R$ and in-plane radius $r = R\sin\theta$, defining Young's contact angle $\theta$. We note immediately from (\ref{dpfin}) that at the top of the drop $\hat{\Pi}$ reduces to $\hat{\Pi}_c=\gamma_{LV}/R$, the Laplace pressure difference across the curved film profile. That is, the pressure under the drop is bigger than outside of it by the well-known amount. If sum rule (\ref{dpfin}) is integrated from outside the fluctuation to the top of the cylindrical cap then one obtains zero, which is an expression of force-balance between the bending of the film profile near the three-phase contact line and the `macroscopic' curvature of the cylindrical cap; i.e. 
\begin{equation}
-\int_{-\infty}^{x_c}dx\left[\hat{\Pi}(x)-\hat{\Pi}_c\right] = r\hat{\Pi}_c = \gamma_{LV}\sin\theta\,\,,
\label{cmsr}
\end{equation}
where $x_c$ denotes the $x$-coordinate of the centre of the cylindrical-cap portion. In the macroscopic limit where the three-phase contact line can be treated as the intersection (at angle $\theta$) of a planar liquid-vapour interface with a planar substrate then Eq.~(\ref{cmsr}) reduces to an integral over the three-phase contact region alone
\begin{equation}
-\int_{-\infty}^{\infty}dx\hat{\Pi}(x) = \gamma_{LV}\sin\theta\,\,,
\label{jrhfb}
\end{equation}
which is the normal component of Young's force balance through the three-phase contact line \cite{adamjrh,unbalanced,derj1,shandeG,jrhpre}. Here, the disjoining pressure is balanced by the distortion of the film surface as one crosses the three-phase contact line. In colloid and interface science this contribution is often equated with the disjoining pressure for planar films (of varying thickness) and is separated out from the macroscopic contribution $\hat{\Pi}_c$; see e.g. \cite{velarde}. When the above integral is taken over the entire surface, in overall planar geometry (Fig.~2), the implied force balance is equivalent to full thermodynamic equilibrium; i.e. $\langle\hat{\Pi}\rangle =0$ as required for thermodynamic equilibrium of a planar film. When the equilibrium geometry is non-planar (e.g. in the presence of a macroscopic adsorbed drop) both (\ref{dpfin}) and (\ref{jrhfb}) average to sum rules of identical form but now involving $\langle\hat{\Pi}\rangle(x)$ and $\langle\hat{\kappa}\rangle(x)$.\\

From the above we can conclude that by construction the interface Hamiltonian theory and associated disjoining pressure is fully consistent with the conservation of momentum perpendicular to the substrate. In field theory, sum rules of this class are known as a Ward identity and are equivalent to the condition of translational invariance (of the entire system) in the $z$-direction \cite{mik-weeks}. It is this physics that enables us to equate the exact disjoining pressure profile with the excess force on the substrate surface underneath the film:
\begin{equation}
\hat{\Pi}[\phi](x) = -\int_{-\infty}^{\infty}dz\left[\rho(x,z;\hat{\ell})-\rho(z;\ell_o)\right]\frac{dv^{ext}_{sub}(z)}{dz}\,\,,
\label{fbgen}
\end{equation}
where the subscript on the external field emphasises that this is the true substrate field without an additional constraint.
Note that (\ref{fbgen}) reduces to (\ref{pism}) when the fluctuations are restricted to planar `breathing modes', in contrast to (\ref{dpfin}). Thus, the general statistical mechanical result (\ref{fbgen}) contains more physics than (\ref{iham}). In principle, sensors could be placed just under the substrate surface to measure the fluctuating disjoining pressure profile $\hat{\Pi}(x)$, provided the fluctuations were sufficiently slow (to enable a local time average before the fluctuation disappeared). It is thus of some interest to note the intimate connection between a measurement of $\hat{\Pi}(x)$ and the associated $\hat{\ell}$. First let us define a local contact angle profile (not to be confused with Young's contact angle):
\begin{eqnarray}
\tan\theta(x) &\equiv& \hat{\ell}'(x)\,\,,\label{tanthx}\\
\sin\theta(x) &\equiv& \frac{\hat{\ell}'}{\sqrt{1+(\hat{\ell}')^2}}\,\,.
\label{sinthx}
\end{eqnarray}
Then the first integral of (\ref{dpfin}) allows us to define $\theta(x)$ in terms of the disjoining pressure profile:
\begin{equation}
\theta(x) = \sin^{-1}\left\{-\frac{1}{\gamma_{LV}}\int_{-\infty}^{x}d\tilde{x}\hat{\Pi}(\tilde{x})\right\}\,\,,
\label{thetasr}
\end{equation}
where the local profile is assumed to be planar at the left boundary. Finally, the integral of (\ref{tanthx}) shows that the local fluctuating profile is fully determined by the disjoining pressure profile defining $\theta(x)$:
\begin{equation}
\hat{\ell}(x) = \int_{-\infty}^{x}d\tilde{x}\tan\theta(\tilde{x})\,\,,
\label{fprof}
\end{equation}
with the zero of film height defined at the far left boundary. In non-planar geometry (such as a system with an equilibrium three-phase contact line) the above results are immediately applicable to the final equilibrium profile \cite{adamjrh}. For overall planar geometry, the true equilibrium film width requires a path integral (\ref{lhat}) over all fluctuations. In principal either of the profiles $\hat{\ell}(x)$ or $\hat{\Pi}(x)$ could be directly measured, but these are not independent measurements because they determine each other through the conservation of momentum normal to the substrate. In a fully molecular system, this link allows one to uniquely define what is meant by the coarse-grained profile $\hat{\ell}(x)$, at least in the absence of overhangs \cite{adamjrh}. One can go further and define correlation function hierarchies from the fluctuating disjoining pressure \cite{mik-weeks}. It has taken the present author his entire working lifetime to grasp the true significance of Derjaguin's disjoining pressure to the statistical mechanics of inhomogeneous fluids.

\section*{Acknowledgments}
I am grateful to A. O. Parry and M. G. Velarde for insightful discussion.


\begin{thebibliography}{9}
\bibitem{footeq} Note that $\gamma_{SG}=(\partial/\partial 1/\ell)(\psi/\ell)$ and that the disjoining pressure is defined to be zero at $\mu=\mu_o$.
\end{thebibliography}
\end{document}